\documentclass[twocolumn,aps,prx,superscriptaddress,pagenumbers,showpacs,helvetica]{revtex4-1}
\usepackage[T1]{fontenc}

\usepackage{amssymb}
\usepackage{amsmath}
\usepackage{graphicx}
\usepackage{array}
\usepackage{multirow}
\usepackage[usenames, dvipsnames]{color}
\usepackage{soul,xcolor}

\newcommand{\bra}[1]{\left< #1 \right\vert}
\newcommand{\ket}[1]{\left\vert #1 \right>}
\newcommand{\pare}[1]{\left( #1 \right)}

\newcommand{\cor}[1]{\left[ #1 \right]}

\newcommand{\llav}[1]{\left\lbrace #1 \right\rbrace}

\newcommand{\PT}{\mathcal{PT}}

\newcommand{\K}{\mathcal{K}}

\begin{document}

\title{Observation of slowly decaying eigenmodes without exceptional points in Floquet, dissipative,  synthetic circuits}
\author{Roberto de J. Le\'on-Montiel}
\thanks{These authors contributed equally.}
\affiliation{Instituto de Ciencias Nucleares, Universidad Nacional Aut\'onoma de
M\'exico,\\ Apartado Postal 70-543, 04510 Cd. Mx., M\'exico}
\email{roberto.leon@nucleares.unam.mx}

\author{Mario A. Quiroz-Ju\'{a}rez}
\thanks{These authors contributed equally.}
\affiliation{Centro de F\'{i}sica Aplicada y Tecnolog\'{i}a Avanzada, Universidad Nacional Aut\'{o}noma de M\'{e}xico, Boulevard Juriquilla 3001, Juriquilla Quer\'{e}taro 76230, M\'{e}xico}
\affiliation{Escuela Superior de Ingenier\'{i}a Mec\'{a}nica y El\'{e}ctrica, Culhuac\'{a}n. Instituto Polit\'{e}cnico Nacional, Santa Ana 1000, San Francisco Culhuac\'{a}''n 04430, Ciudad de M\'{e}xico, M\'{e}xico}

\author{Jorge L. Dom\'{i}nguez-Ju\'{a}rez}
\thanks{These authors contributed equally.}
\affiliation{Centro de F\'{i}sica Aplicada y Tecnolog\'{i}a Avanzada, Universidad Nacional Aut\'{o}noma de M\'{e}xico, Boulevard Juriquilla 3001, Juriquilla Quer\'{e}taro 76230, M\'{e}xico}
\affiliation{C\'{a}tedras CONACyT, CFATA, Universidad Nacional Aut\'{o}noma de M\'{e}xico, Juriquilla, Quer\'{e}taro 76230, M\'{e}xico.}

\author{Rafael Quintero-Torres}
\thanks{These authors contributed equally.}
\affiliation{Centro de F\'{i}sica Aplicada y Tecnolog\'{i}a Avanzada, Universidad Nacional Aut\'{o}noma de M\'{e}xico, Boulevard Juriquilla 3001, Juriquilla Quer\'{e}taro 76230, M\'{e}xico}

\author{Jos\'{e} L. Arag\'{o}n}
\affiliation{Centro de F\'{i}sica Aplicada y Tecnolog\'{i}a Avanzada, Universidad Nacional Aut\'{o}noma de M\'{e}xico, Boulevard Juriquilla 3001, Juriquilla Quer\'{e}taro 76230, M\'{e}xico}

\author{Andrew K. Harter}
\affiliation{Department of Physics, Indiana University - Purdue University Indianapolis (IUPUI), Indianapolis, Indiana 46202 USA}
\email{yojoglek@iupui.edu}

\author{Yogesh N. Joglekar}
\affiliation{Department of Physics, Indiana University - Purdue University Indianapolis (IUPUI), Indianapolis, Indiana 46202 USA}

\setstcolor{red}

\date{\today}

\begin{abstract}
We report the first experimental observation of multiple transitions showing the emergence and disappearance of slowly decaying eigenmodes in a dissipative, Floquet electronic system with synthetic components. Conventional wisdom has it that such transitions occur at exceptional points, and avoided-level-crossing driven phenomena in purely dissipative systems are formerly unexplored. Remarkably, in our system, the slowly decaying eigenmodes emerge without exceptional points. Our experimental setup makes use of an LC oscillator inductively coupled to an RLC oscillator, where the time-periodic (Floquet) inductive coupling and resistive-heating losses can be independently controlled by means of external voltage signals. With a Floquet dissipation, we observe that slowly-decaying eigenmodes emerge at vanishingly small dissipation strength in the weak coupling limit. With a moderate, Floquet coupling, multiple instances of their emergence and disappearance are observed. With an asymmetric dimer model, we show that these transitions, driven by avoided-level-crossing in purely dissipative systems, are generically present in both static and Floquet domains.
\end{abstract}

\maketitle

\section{Introduction}
\label{sec:intro}
Over the past five years, systems described by non-Hermitian, parity-time ($\mathcal{PT}$) symmetric Hamiltonians have become a subject of intense research~\cite{feng2017,rami2018}. Such a Hamiltonian $H_{PT}$ is invariant under the combined operations of parity ($\mathcal{P}$) and time-reversal ($\mathcal{T}$), but it does not commute with either of the two. When its non-Hermiticity is small, its eigenvalues $\lambda_k$ are purely real, i.e. $\Im\lambda_k=0$, and its eigenvectors are simultaneous eigenvectors of the antilinear $\mathcal{PT}$ operator with eigenvalue one. The spectrum changes into complex-conjugate pairs when the non-Hermiticity exceeds a threshold called the $\mathcal{PT}$ symmetry breaking threshold~\cite{bender1998,review2013}. At the threshold, two or more eigenvalues of $H_{PT}$ become degenerate as do the corresponding eigenvectors, i.e. the $\mathcal{PT}$ symmetry breaking point is an exceptional point (EP) of the Hamiltonian~\cite{kato}. While not fundamental in their origin~\cite{bender2004,mostafa,violate}, $\PT$-symmetric Hamiltonians faithfully describe open classical systems with balanced, spatially separated, gain and loss, and have been experimentally realized in photonic lattices~\cite{ganainy2007,makris2008,ruter2010,regensburger2012} , microring resonators~\cite{peng2014,hodaei2014,peng2014-2}, superconducting wires~\cite{chtchelkatchev2012}, and electrical circuits~\cite{schindler2011,chitsazi2017}. In addition to studying the dynamics across the $\mathcal{PT}$ transition, experiments on these classical systems have observed the enhanced sensitivity near~\cite{ep3,ep2} and topological properties~\cite{dynamicEP} of the  EP at the $\mathcal{PT}$-symmetry breaking threshold.

When a system has unbalanced gain and loss, the eigenvalues of its non-$\mathcal{PT}$-symmetric Hamiltonian are, in general, complex and the exceptional point is replaced by an avoided level crossing (ALC)~\cite{weissALC}. ALC refers to the flow of complex, non-degenerate eigenvalues towards, and then away from, each other~\cite{rotterALC,rotterALC2}. A laser, with its constant cavity loss and a pump-current adjustable gain, is a prototypical system with local, unbalanced gain and loss. In such coupled lasers, many counter-intuitive transitions, such as pump-induced laser death~\cite{rotter2012,rotter2014nc}, loss-induced suppression and revival of lasing~\cite{rotter2014sci}, and laser self-termination~\cite{lstpra,lsttrimer} have been observed, or predicted, based on the ALC. In all cases, however, the transitions occur at parameters when the maximum of the imaginary part of complex  eigenvalues, $\max\Im\lambda_k$, changes sign, and system parameters where the ALC occurs, i.e the distance between the eigenvalues in the complex plane is the shortest, do not signal any transition.

Apart from a shift along the imaginary axis, the Hamiltonian for a two-level, neutral-loss system is the same as that of a two-level, gain-loss system. Based on this observation, the language of $\mathcal{PT}$ symmetry has been adopted to systems with localized dissipation that are ``identity-shifted'' from a $\mathcal{PT}$ symmetric Hamiltonian~\cite{ornigotti2014}. When the loss strength $\gamma$ is small, the two eigenmodes of the dissipative Hamiltonian $H_D$ have the same decay rate, both of which increase with $\gamma$. The passive $\mathcal{PT}$ transition is signaled by the emergence of two different decay rates, one of which decreases as $\gamma$ is increased beyond the threshold. Indeed, the first observation of a $\mathcal{PT}$ transition was in a waveguide dimer, one with loss and the other without loss~\cite{Guo2009}, wherein the net transmission {\it increased} with increase in the local loss,  due to the emergence of the slowly decaying mode. In the strictest sense, the Hamiltonian $H_D$ does not commute with the $\mathcal{PT}$ operator and thus eigenvectors of $H_D$ cannot be simultaneous eigenvectors of the $\mathcal{PT}$ operator with eigenvalue unity.  However, based on earlier observations regarding equal decay rates vs. emergence of a slow mode, there are two ways to define a passive $\mathcal{PT}$-symmetry breaking transition.

For dissipative Hamiltonians that are identity-shifted from a balanced gain-loss Hamiltonian $H_{PT}$, the transition is defined as a transition where the spectrum of $H_D$ changes from all eigenmodes having identical decay rates to different decay rates. A second -- physically transparent -- way to define the passive $\mathcal{PT}$ transition is as follows. When the local loss $\gamma$ is small, the eigenmode decay rates $\Gamma_k\equiv -\Im\lambda_k>0$ increase with $\gamma$. Past a critical value $\gamma_{PT}$, a slowly decaying mode emerges, whose decay rate $\Gamma_s(\gamma)$ decreases when the loss is increased further. The passive $\mathcal{PT}$-symmetry breaking threshold $\gamma_{PT}$, then, is defined by~\cite{caveat}
\begin{equation}
\label{eq:passiveptb}
d\Gamma_s(\gamma_{PT}^{-})/d\gamma>0\,\,\mathrm{and}\,\,d\Gamma_s(\gamma_{PT}^{+})/d\gamma<0.
\end{equation}
The second definition encompasses dissipative Hamiltonians that are not identity shifted from a balanced gain-loss Hamiltonian and therefore do not have an EP at the threshold $\gamma_{PT}$. The emergence of a slowly decaying mode -- the key, experimentally observed signature~\cite{Guo2009,zeuner2015,weimann2017} -- does not depend upon whether it occurs at an exceptional point or not. Therefore, in this paper, we call the regions of the parameter-space where the eigenmode decay rates, equal or not, increase with $\gamma$ as the ``passive $\mathcal{PT}$-symmetric regions''. Similarly, we call regions where the slowly decaying eigenmodes exist, arising from an identity-shifted $H_{PT}$ or not, as ``passive $\mathcal{PT}$-symmetry broken'' regions.

There is a compelling reason for studying the systems with localized dissipation. The fundamental obstacle to a quantum system with Hamiltonian $H_{PT}$ is that amplification is accompanied by quantum noise~\cite{haus1962,cava1982}. In optical settings, at a few-photon level, the gain is randomized by spontaneous emission, while the loss is statistically linear down to a single-photon level. Consequently, there are no experimental realizations of gain-loss systems that show quantum correlations present. A dissipative system, on the other hand, can be implemented down to the quantum level. Equation~(\ref{eq:passiveptb}) provides a clear, physically intuitive way forward to define the passive $\mathcal{PT}$ transition in truly quantum systems which may or may not be identity-shifted from a $\mathcal{PT}$ symmetric Hamiltonian $H_{PT}$, and the two definitions are equivalent for identity-shifted dissipative Hamiltonians with EPs. This approach has led to the first observation of passive $\mathcal{PT}$ breaking transitions in the quantum domain with correlated single photons~\cite{xiao2017} and a non-interacting, degenerate Fermi gas~\cite{Li2016}. The emergence of a slowly decaying mode in purely dissipative systems without EPs and its relation to the ALC have been formerly unexplored.

In this paper, we theoretically and experimentally demonstrate that inductively coupled electronic circuits have the passive $\mathcal{PT}$ symmetry breaking transition in the absence of EPs in both static and Floquet domains, Our system comprises a neutral LC oscillator inductively coupled to an RLC oscillator with Joule-heating loss (Sec.~\ref{sec:static}). With static and time-periodic (Floquet) versions of the effective, lossy Hamiltonian that describes this system, we characterize the ``passive $\mathcal{PT}$-symmetric'' and  ``passive $\mathcal{PT}$-symmetry broken'' regions, and observe the emergence of slowly decaying eigenmodes that are indicative of the passive $\mathcal{PT}$ broken region (Secs.~\ref{sec:static} and~\ref{sec:floquetloss}). Experimental results for the circuit energy dynamics in the moderate Floquet coupling regime, showing multiple ``passive $\mathcal{PT}$ symmetry breaking'' transitions are presented in Sec.~\ref{sec:floquetLx}. In Sec.~\ref{sec:disc}, by using an asymmetrical dimer model, we show that passive $\mathcal{PT}$ transitions at the ALC occur in both static and Floquet dissipative Hamiltonians. In contrast to the experiments on coupled LC circuits with gain and loss~\cite{schindler2011,chitsazi2017}, our system undergoes passive $\mathcal{PT}$-symmetry breaking transitions without EPs, and therefore signals a new, ALC-driven paradigm that is applicable to a broad array of purely dissipative classical and quantum systems.



\begin{figure*}
\centering
\includegraphics[width=\textwidth]{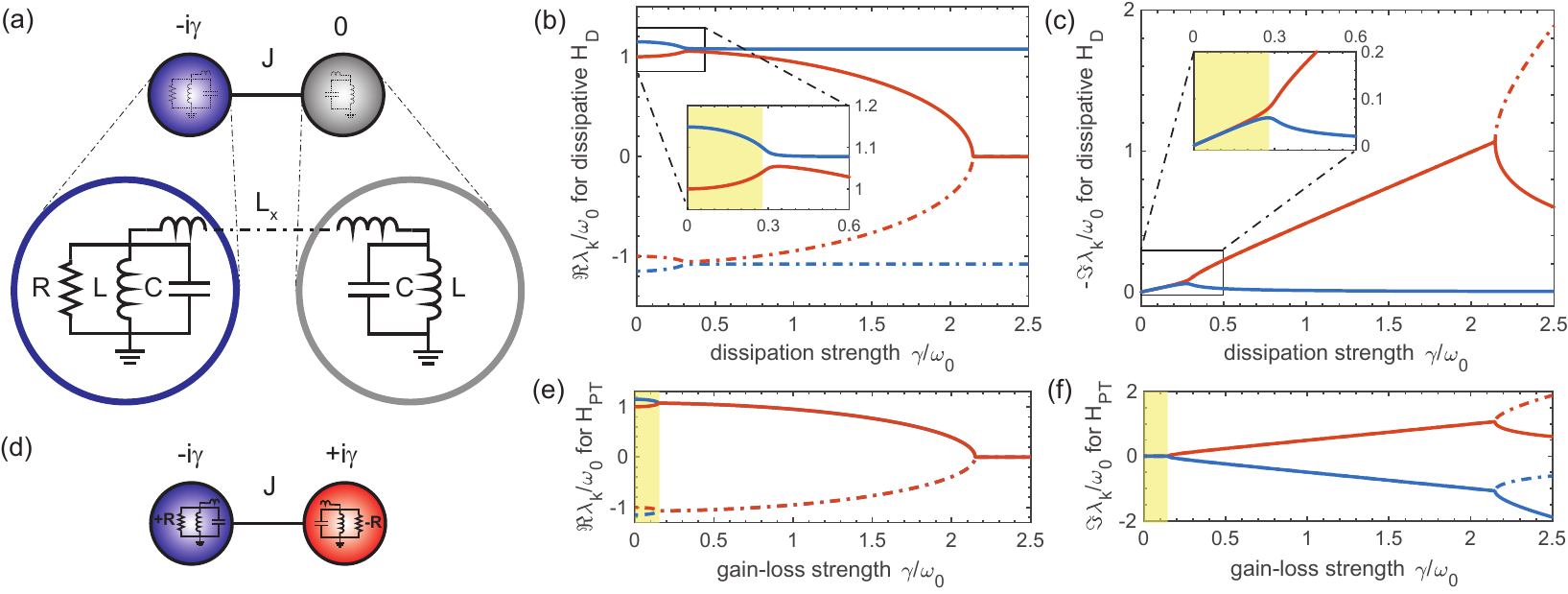}
\caption{Passive $\mathcal{PT}$ symmetry breaking transition without an EP. (a) schematic of an LC oscillator (gray) inductively coupled to an RLC oscillator (blue). In the weak coupling limit, this system maps onto a dissipative dimer (Methods, section D). (b) flow of eigenvalues $\Re\lambda_k$ as a function of loss shows that the top two (red and blue) levels attract each other and reach a minimum gap $\propto M^3$ near $\gamma/\omega_0\sim 2M^2$ before diverging again, thus indicating an ALC. (c) flow of decay rates $-\Im\lambda_k$ shows that slowly decaying modes (blue) emerge at $\gamma/\omega_0\sim 2M^2$ signaling a passive $\mathcal{PT}$ transition at the location of the ALC. The insets show expanded view of the transition region. (d) schematic circuit with gain and loss~\cite{schindler2011}. (e)-(f) flows of $\Re\lambda_k$ and $\Im\lambda_k$ as a function of the gain-loss strength show that the $\PT$-breaking transition occurs only an exceptional point, $\gamma=(\omega_M-\omega_0)\propto \omega_0M^2$. The comparison between (a) and (d) shows that the transition accompanied by an ALC, instead of an EP, occurs only in the dissipative case.}
\label{fig:static}
\end{figure*}

\section{Theory: passive $\mathcal{PT}$-transition in a static Hamiltonian with ALC}
\label{sec:static}
Our system is a neutral LC oscillator, formed by a synthetic inductor and capacitor, inductively coupled to another parallel, synthetic RLC circuit with a coupling inductor $L_x$ (see Fig.~\ref{fig:static}a). In parity-time symmetric systems one generally investigates the dynamics of the local density of a quantity $\mathcal{Q}(t)\equiv\langle\psi(t)|\psi(t)\rangle$ that is conserved when the system is isolated and described by a Hermitian Hamiltonian. For optical $\mathcal{PT}$ systems, $\mathcal{Q}$ is the total energy in the electromagnetic pulse or, equivalently, the number of photons, and $|\psi(t)\rangle$ then represents the location-dependent amplitude of the electric-field envelope; in a passive $\mathcal{PT}$ system with ultracold atoms, $\mathcal{Q}(t)$ is the total number of atoms in the trap and $|\psi(t)\rangle$ is the hyperfine-level associated wavefunction of a single atom. In our case, the time-dependent energy stored in and across the two coupled oscillators is given by a positive-definite quadratic form, i.e. $\mathcal{Q}(t)\equiv\langle\psi(t)|\psi(t)\rangle=\langle\phi(t)|A|\phi(t)\rangle$. Here $A=\mathrm{diag}(C,C,L,L,L_x)/2$ is a real, diagonal matrix, and $|\phi\rangle=(V_1,V_2,I_1,I_2,I_x)^T$ is a real column-vector comprising the voltages $V_{1,2}$ across the two capacitors, the currents $I_{1,2}$ across the two inductors, and the current $I_x$ flowing across the coupling inductor. The decay dynamics of the energy $\mathcal{Q}(t)$ in the system is determined by Kirchoff laws and leads to a Schrodinger-like equation $i\partial_t|\psi(t)\rangle = H_D|\psi(t)\rangle$ (Methods, section A). The rank-4, $5\times 5$ lossy Hamiltonian is given by
\begin{equation}
H_D(\gamma)= \left[
\begin{array}{ccccc}
-i\gamma & 0 & -i\omega_0 & 0 & -i\omega_0 M\\
0 & 0 & 0 & -i\omega_0 & i\omega_0 M\\
i\omega_0 & 0 & 0 & 0 & 0 \\
0 & i\omega_0 & 0 & 0 & 0 \\
i\omega_0 M& -i\omega_0 M & 0 & 0 & 0\\
\end{array} \right].
\label{eq:Hd}
\end{equation}
Here, $\omega_0=1/\sqrt{LC}$ is the frequency of an isolated oscillator, $M=\sqrt{L/L_x}$ is the dimensionless coupling between the two oscillators, and $\gamma=1/RC$ is the dissipation rate of the parallel RLC oscillator. Because this is a classical system, $H_D$ has purely imaginary entries;  it ensures that the ``state vector'' $|\psi(t)\rangle$ remains real at all times. Apart from the trivial eigenvalue $\lambda=0$, the characteristic equation for $H_D$ is given by
\begin{equation}
\label{eq:eeloss}
(\lambda^2-\omega_0^2)(\lambda^2-\omega_M^2)-i\gamma\lambda\left[\lambda^2-\omega_0^2(1+M^2)\right]=0,
\end{equation}
where $\omega_M=\omega_0\sqrt{1+2M^2}$. It follows from Eq.(\ref{eq:eeloss}) that if $\lambda$ is an eigenvalue of the Hamiltonian $H_D$, so is $-\lambda^{*}$.

The resulting flow of eigenvalues $\Re\lambda_k(\gamma)$ and $\Im\lambda_k(\gamma)$ as a function of the dissipation, for $M=0.4$, are shown in Fig.~\ref{fig:static}b and Fig.~\ref{fig:static}c respectively. Starting from $\pm\omega_0$ (red solid and dot-dashed lines) and $\pm\omega_M$ (blue solid and dot-dashed lines), the $\Re\lambda_k$ approach each other as $\gamma/\omega_0$ is increased. The levels reach a minimum gap $\propto M^3$ at loss strength $\gamma/\omega_0\propto 2M^2$ and then they diverge again, i.e. the static Hamiltonian has an ALC near $\gamma\sim 2M^2\omega_0$. Note that due to this scaling, in the weak coupling limit $M\ll 1$, the gap appears to vanish, just as it does in the balanced gain-loss electrical circuits (Fig.~\ref{fig:static}e)~\cite{schindler2011}. The inset shows an enlarged view of the transition region. Figure~\ref{fig:static}c shows the evolution of the doubly-degenerate decay rates $\Gamma_k\equiv-\Im\lambda_k$. At small dissipation, both (red and blue) decay rates increase with $\gamma$. However, at $\gamma/\omega_0\sim 2M^2$,  two ``slowly decaying'' (blue) eigenmodes with $d\Gamma_s/d\gamma<0$ emerge, indicating a passive $\mathcal{PT}$ symmetry breaking transition which occurs at the location of the ALC. The inset shows an enlarged view of this region, where the shaded part indicates ``passive $\mathcal{PT}$ symmetric'' region ($d\Gamma_s/d\gamma>0$) and the unshaded part indicates the ``passive $\mathcal{PT}$-symmetry broken'' region ($d\Gamma_s/d\gamma<0$). In contrast, we note that in systems with unbalanced gain and loss~\cite{lstpra,lsttrimer}, the interesting physical phenomena~\cite{rotter2014nc,rotter2014sci} do not occur at the location of the ALC, but instead at system parameters where $\max\Im\lambda_k$ changes sign.

The results for an RLC oscillator coupled to a gain-LC oscillator, Fig.~\ref{fig:static}d, are shown in the subsequent panels~\cite{schindler2011}. Note that the corresponding Hamiltonian $H_{PT}(\gamma)$ is identical to $H_D(\gamma)$ except for an additional nonzero term given by $H_{PT}(2,2)=+i\gamma$. The resulting flow of $\Re\lambda_k$, Fig.~\ref{fig:static}e, shows that starting from $\pm\omega_0$ (red solid and dot-dashed lines) and $\pm\omega_M$ (blue solid and dot-dashed lines), the levels for $\Re\lambda_k$ attract each other and become degenerate at the exceptional point $\gamma=\omega_M-\omega_0$. Fig.~\ref{fig:static}f shows that starting from zero, $\Im\lambda_k$ take off in a characteristic square-root pattern at the same EP, i.e $\gamma=(\omega_M-\omega_0)\approx\omega_0M^2$. As an aside, we note that both Hamiltonians  have an exceptional point deep in the $\mathcal{PT}$ symmetry broken region, at $\gamma/\omega_0\sim 2\sqrt{1+M^2}$. However, in either case, this EP does not signal any transition.

The results in Fig.~\ref{fig:static} predict that in our lossy, static system, a slowly decaying eigenmode emerges at the location of the ALC.  In two-level systems, when a static Hamiltonian is replaced by its Floquet version, a rich phase diagram with multiple $\mathcal{PT}$ symmetric and $\mathcal{PT}$ symmetry broken regions separated by lines of EPs emerges~\cite{joglekar2014,lee2015}. In particular, the system has a $\mathcal{PT}$ broken region at arbitrarily small $\gamma$ if the loss-modulation frequency matches the energy gap~\cite{Li2016}. Armed with these insights, we now experimentally investigate the fate of the slowly decaying eigenmodes in this system in the presence of periodic loss or coupling.


\section{Experimental Results}

\subsection{Passive $\mathcal{PT}$ transitions with Floquet dissipation}
\label{sec:floquetloss}

\begin{figure*}
\centering
 \includegraphics[width=\textwidth]{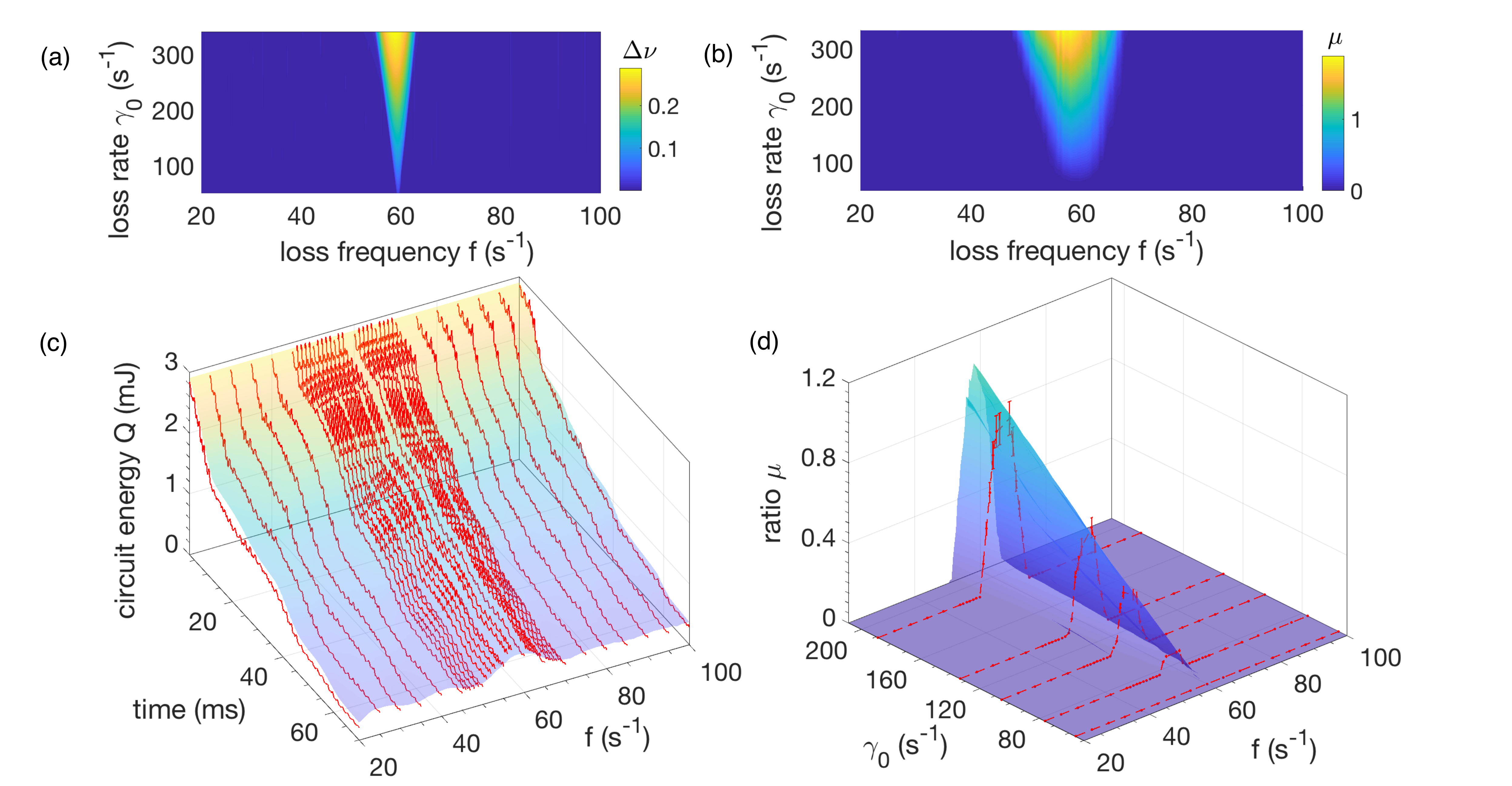}
\caption{Observation of Floquet-dissipation induced slowly-decaying eigenmode in a system without EPs. (a) the phase diagram in the $(\gamma_0,f)$ plane shows that the passive $\mathcal{PT}$ symmetry broken phase ($\Delta\nu>0$) occurs at vanishingly small $\gamma_0$ in the vicinity of $f=60$ s$^{-1}$. (b) experimentally friendly ratio $\mu$, Eq.(\ref{eq:ratio}), shows the same qualitative features, although $\mu>0$ region is broadened due to a finite $\tau$ value. (c) experimentally measured circuit energy $\mathcal{Q}(t,f)$ traces (red lines) show a clear slowdown of the decay, indicative of the emergence of a slowly decaying eigenmode, in the vicinity of $f=60$ s$^{-1}$ and match well with theoretical predictions (surface plot). The loss strength $\gamma_0=77$ s$^{-1}$ is far smaller than the loss strength $\gamma\sim 2M^2\omega_0$ necessary in the static case. (d) the ratio $\mu(\gamma_0,f)$, obtained from experimental data, shows that the system goes from passive $\mathcal{PT}$ symmetric region, to passive $\mathcal{PT}$-symmetry broken region, to passive $\mathcal{PT}$ symmetric region as the loss-modulation frequency $f=1/T_f$ is swept (red: data; surface plot: theory).}
\label{fig:floquetloss}
\end{figure*}

We implement a circuit where the resistance in the lossy (RLC) unit is switched between an open circuit and $R_0$ during a time period $T_f\equiv 1/f$. The time-periodic, dissipative Hamiltonian in this case is given by Eq.(\ref{eq:Hd}) with a square-wave dissipation function, i.e.,
\begin{equation}
\label{eq:fl}
\gamma(t)=\left\{\begin{array}{cc}
\gamma_0 & 0\leq t\leq T_f/4,\\
0 & T_f/4 \leq t\leq 3T_f/4,\\
\gamma_0 & 3T_f/4\leq t\leq T_f.
\end{array}\right.
\end{equation}
The lossy Hamiltonian $H_D$ is shifted along the imaginary axis from a $\mathcal{PT}$ symmetric Hamiltonian $H_{PT}$ by a {\it non-identity, diagonal matrix} ${\bf I}_2=\mathrm{diag}(1,1,0,0,0)$, i.e
\begin{equation}
H_D(\gamma)=H_{PT}(\gamma/2)-i\frac{\gamma}{2}{\bf I}_2.
\end{equation}
Since ${\bf I}_2$ is not invariant under arbitrary, change-of-basis transformations, $H_D(t)$ and $H_{PT}(t)$ do not share the same topological structure for their static or Floquet eigenvalue spectra. The $\mathcal{PT}$ phases of the Floquet Hamiltonian $H_D(t)$ are determined by the eigenvalues $\nu_k$ of the one-period time-evolution operator $G_D(T_f)={\bf T}\exp\left[-i\int_0^{T_f} dt' H_D(t')\right]$ where ${\bf T}$ stands for the time-ordered exponential~\cite{monodromy1,monodromy2}. Because we have a piecewise constant Hamiltonian, Eq.(\ref{eq:fl}), the monodromy matix $G_D(T_f)$ can be explicitly calculated. In addition to the trivial eigenvalue $\nu=1$, which reflects the rank-4 nature of the $5\times 5$ Hamiltonian $H_D(t)$, the remaining eigenvalues $\nu_k$ of $G_D(T_f)$ give four dissipative quasienergies $\lambda_k\equiv\ln\nu_k$ that also occur in pairs $(\lambda,-\lambda^{*})$. Thus, there are two distinct, particle-hole symmetric, frequency values $|\Re\lambda_k|$ and two decay rates $-\Im\lambda_k>0$ for our system. The passive $\mathcal{PT}$-symmetric phase is signaled by $\Delta\nu\equiv(\max|\nu_k|-\min|\nu_k|)\sim 0$ and $\Delta\nu>0$ indicates a passive $\mathcal{PT}$-symmetry broken phase~\cite{obusePT}. However, due to the presence of two frequencies and two decay rates that have to be determined from the decaying voltage and current signals, this approach is not experimentally suitable.

An alternate, experimentally friendly approach to track the passive, $\mathcal{PT}$ symmetry breaking transition is to define a scaled energy,
\begin{equation}
\label{eq:scaledenergy}
E(t)=\langle\psi(0)| G^{\dagger}_D(t) e^{+\gamma{\bf I}_2t}G_D(t)|\psi(0)\rangle.
\end{equation}
For a dissipative two-level system, this scaled quantity shows oscillatory behavior in the $\mathcal{PT}$ symmetric phase, with its amplitude and period both diverging as the system approaches the $\mathcal{PT}$ phase boundary, and an exponential rise with time in the $\mathcal{PT}$ symmetry broken phase~\cite{ornigotti2014,xiao2017,Li2016}. This qualitative difference is quantified by the ratio
\begin{equation}
\label{eq:ratio}
\mu=\text{log}\llav{\frac{\text{max}\cor{E\pare{0\leq t\leq 2\tau}}}{\text{max}\cor{E\pare{0\leq t\leq \tau}}}},
\end{equation}
where $\tau$ is an arbitrary (large) time window. When $\mu=0$, the system is in the passive $\PT$-symmetric phase, while $\mu > 0$ reveals the rate of exponential growth of the scaled energy in the passive $\PT$-broken phase. This procedure provides an operationally straightforward metric to track the transitions between the passive $\mathcal{PT}$-symmetric and passive $\mathcal{PT}$-symmetry broken regions in the two-dimensional parameter space $(\gamma_0,f)$ of Floquet dissipation.

We experimentally implement the system described in Eqs.(\ref{eq:Hd}) and (\ref{eq:fl}) by using functional blocks synthesized with operational amplifiers and passive linear electrical components. (See Methods, section B, and Refs.~\cite{leon2015,quiroz2016} for details.) Our experimental setup is designed so that the synthetic inductance and capacitance in each oscillator are $L$=1 mH and $C$=0.1 mF respectively, leading to natural frequency of each oscillator $\omega_0/2\pi$=503 s$^{-1}$. The remaining parameters of the electronic circuit are defined depending on the specific configuration of the system. For the dynamic-dissipation case, the coupling inductor is set to $L_{x}$=8 mH ($M=1/2\sqrt{2}=0.35$) and the resistance is periodically driven by means of an external square-wave signal. The maximum value of the resistance is $R_{\max}$=400 $\Omega$, i.e. $\min\gamma(t)$=25 s$^{-1}$. The minimum resistance is selected from $R_{\min}$=\{50, 75, 95, 130, 180\} $\Omega$, and gives $\gamma_0$=\{200, 133, 105, 77, 56\} s$^{-1}$ respectively. The parallel resistance in the neutral LC circuit is $R_N$=1 k$\Omega$ and leads to a loss-rate $\gamma_N$=10 s$^{-1}$ that is far smaller than the loss rate $\gamma(t)$ in the RLC circuit. In all cases, our time-trace data are take up to $t_{\max}$=200 ms, beyond which the effects of the resistor $R_N$ become relevant.

Figure~\ref{fig:floquetloss}a shows that the numerically obtained phase diagram $\Delta\nu(\gamma_0,f)$ has a triangular passive $\mathcal{PT}$-symmetry broken region centered at $f=60$ s$^{-1}$. In its neighborhood, the system is driven from a passive $\mathcal{PT}$-symmetric phase to the passive $\mathcal{PT}$-symmetry broken phase and back at vanishingly small loss-strength by sweeping the frequency $f$ of the Floquet dissipation~\cite{joglekar2014,lee2015}. Figure~\ref{fig:floquetloss}b shows that the experimentally friendly ratio $\mu(\gamma_0,f)$, obtained by numerically solving the Kirchoff-law differential equations (Methods, section A) and using $\tau$=40 ms, has the same features. Because of the divergent period of $E(t)$ oscillations near the phase boundary, at a finite $\tau$, points in the passive $\mathcal{PT}$-symmetric regions with period $\gtrsim\tau$ also exhibit a positive ratio $\mu$, and broaden the $\mu>0$ region in Fig.~\ref{fig:floquetloss}b compared to the $\Delta\nu>0$ region in Fig.~\ref{fig:floquetloss}a. In both cases, the loss-strength $\gamma_0$ is one order of magnitude smaller than the static passive $\mathcal{PT}$-symmetry breaking threshold $\sim2M^2\omega_0$.

Figure~\ref{fig:floquetloss}c shows the experimentally measured time-traces for the circuit energy $\mathcal{Q}(t)$ obtained for $\gamma_0=77$ s$^{-1}$ and different loss-modulation frequencies $f$ (red lines: data; surface plot: theory). As the modulation frequency is changed from 40 s$^{-1}$ to 60 s$^{-1}$, the decay rate for $\mathcal{Q}(t)$ dramatically slows down and signals the emergence of a  slowly decaying mode, i.e. the passive $\mathcal{PT}$-symmetry broken phase. Increasing the modulation frequency further to 80 s$^{-1}$ drives the system back into the passive $\mathcal{PT}$ symmetric phase. (See Methods, section C, for experimental time-traces with additional values of $\gamma_0$.) Figure~\ref{fig:floquetloss}d shows that $\mu(\gamma_0,f)$, obtained from Eq.(\ref{eq:ratio}) with $\tau$=40 ms, changes from zero to maximum as $f$ is swept from 40 s$^{-1}$ to 60 s$^{-1}$, and drops back to zero when $f$ is increased further to 80 s$^{-1}$. The red points are data (with 5\% error bars); the surface plot is from theory. The frequency-averaged, time-integrated relative error between theory and experimental results in Fig.~\ref{fig:floquetloss}c is $\delta{\mathcal Q}=-0.025\pm0.076$ and in Fig.~\ref{fig:floquetloss}d is $\delta\mu=0.0021\pm0.0017$ (Methods, section C).

\subsection{Circuit energy dynamics with Floquet coupling}
\label{sec:floquetLx}

\begin{figure*}
\centering
\includegraphics[width=\textwidth]{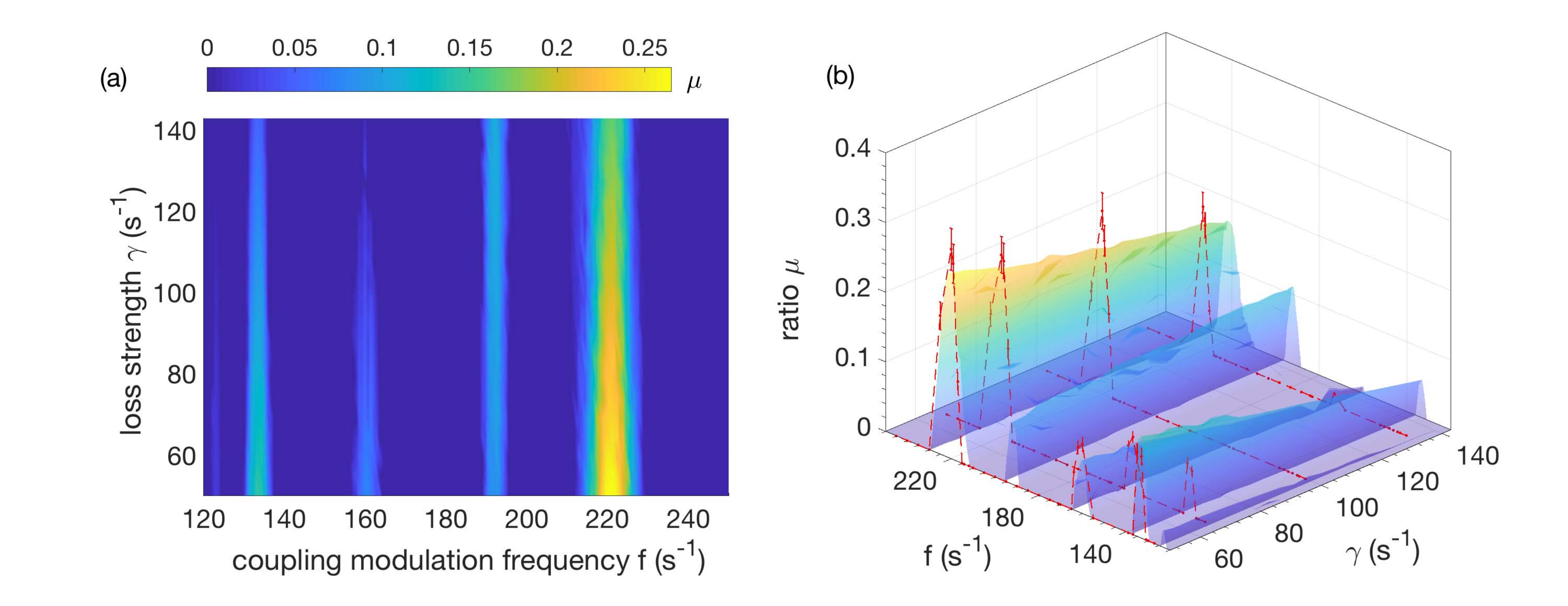}
\caption{Circuit energy $\mathcal{Q}(t)$ decay dynamics with moderate, Floquet coupling. (a) numerically obtained ratio $\mu(\gamma,f)$, Eq.(\ref{eq:ratio}), shows the emergence of slowly decaying modes ($\mu>0$) at loss strength $\gamma/\omega_0\ll 1$, most prominently at $f=220$ s$^{-1}$. (b) experimentally obtained $\mu$-values capture this mode at $f=220$ s$^{-1}$, and partially capture other, weaker cases of decay slowdown with $f\sim$ \{130, 160\} s$^{-1}$ (red: data, surface: theory). The dimensionless coupling is varied between $M=0.707$ and $M=0.5$, and a time-window of $\tau=7$ ms is used. At longer times $t\gtrsim 2\tau$, instabilities and parasitic losses make the Floquet coupling data unreliable.}
\label{fig:floquetLx}
\end{figure*}

In this subsection, we experimentally explore the dynamics of circuit energy $\mathcal{Q}(t)$ when the coupling $M(t)$ is periodically varied. The effective Hamiltonian in this case is given by Eq.(\ref{eq:hdlx}). In addition to the constant dissipative term, it has a periodic driving term $-i\partial_t\ln M(t)$ that, on average, does not add or subtract energy from the system. We investigate the emergence of a slowly-decaying eigenmode in this system by tracking the circuit energy $\mathcal{Q}(t,f)$ and the ratio $\mu(\gamma_0,f)$, when the coupling inductance $L_x$ is varied from $L_{\min}$=2 mH ($M=0.707$) to $L_{\max}$=4 mH ($M=0.5$) in a square-wave fashion over a period $T_f=1/f$. The values for the static resistance in this configuration are $R$=\{75, 100, 150, 200\} $\Omega$, and correspond to loss rates $\gamma$=\{133, 100, 67, 50\} s$^{-1}$.

Figure~\ref{fig:floquetLx}a shows the numerically obtained phase diagram for the ratio $\mu(\gamma,f)$, where $\mu\sim 0$ indicates a regime where the circuit energy $\mathcal{Q}(t)$ decays rapidly and the eigenmode decay rates increase $\gamma$ (passive $\mathcal{PT}$-symmetric region). In contrast, the regions with $\mu>0$ denote emergence of a slowly decaying eigenmode (passive $\mathcal{PT}$-symmetry broken region). The experimentally obtained values of the ratio $\mu$ are compared with the theoretical predictions in Fig.~\ref{fig:floquetLx}b (red: data, surface: theory). We see that the emergence of the slowly decaying mode near $f=220$ s$^{-1}$ is clearly visible in the data, whereas the other, weaker, peaks are only partially captured. The frequency-averaged, time-integrated relative error between theory and experimental results in Fig.~\ref{fig:floquetLx}b is $\delta{\mathcal Q}=-0.038\pm0.071$ and $\delta\mu=0.0076\pm0.027$ (Methods, section C). The larger error in the Floquet coupling case is a consequence of the instabilities produced by the injection (removal) of energy into (from) the system, which is produced by the periodic changes of the coupling inductance, Eq.(\ref{eq:hdlx}). These instabilities and resulting parasitic losses become increasingly dominant after half-a-dozen Floquet cycles, and thus limit the time range for reliable data to $2\tau\sim$ 15 ms.

We note that the multiple emergences of slowly decaying eigenmodes over a small range of coupling modulation frequency is a salient feature of the not-weakly-coupled oscillators. In the weak coupling limit $M\ll 1$, periodic variations of $L_x$ translate into square-wave variation of the effective dimer coupling $J\sim M^2\omega_0/2$ (Methods, section C). Such Floquet dimer coupling leads to passive $\mathcal{PT}$-symmetry broken regions at vanishingly small dissipation strength $\gamma_0$ only in the neighborhood of resonances $2\pi f/J=1,1/2,1/3,\ldots$~\cite{Li2016}. In contrast, results in Fig.~\ref{fig:floquetLx} demonstrate emergence of slowly decaying eigenmodes at frequencies that are far off the resonance values.


\section{Discussion}
\label{sec:disc}
In this paper, we have presented the theory and experimental observation of passive $\mathcal{PT}$ symmetry breaking transitions, driven by avoided level crossing, in a dissipative, synthetic circuit with static and time-periodic parameters. We have observed multiple instances of the emergence of slowly decaying eigenmodes at loss strengths that are one order of magnitude smaller than the static threshold loss strength.


\begin{figure*}
\centering
\includegraphics[width=0.95\textwidth]{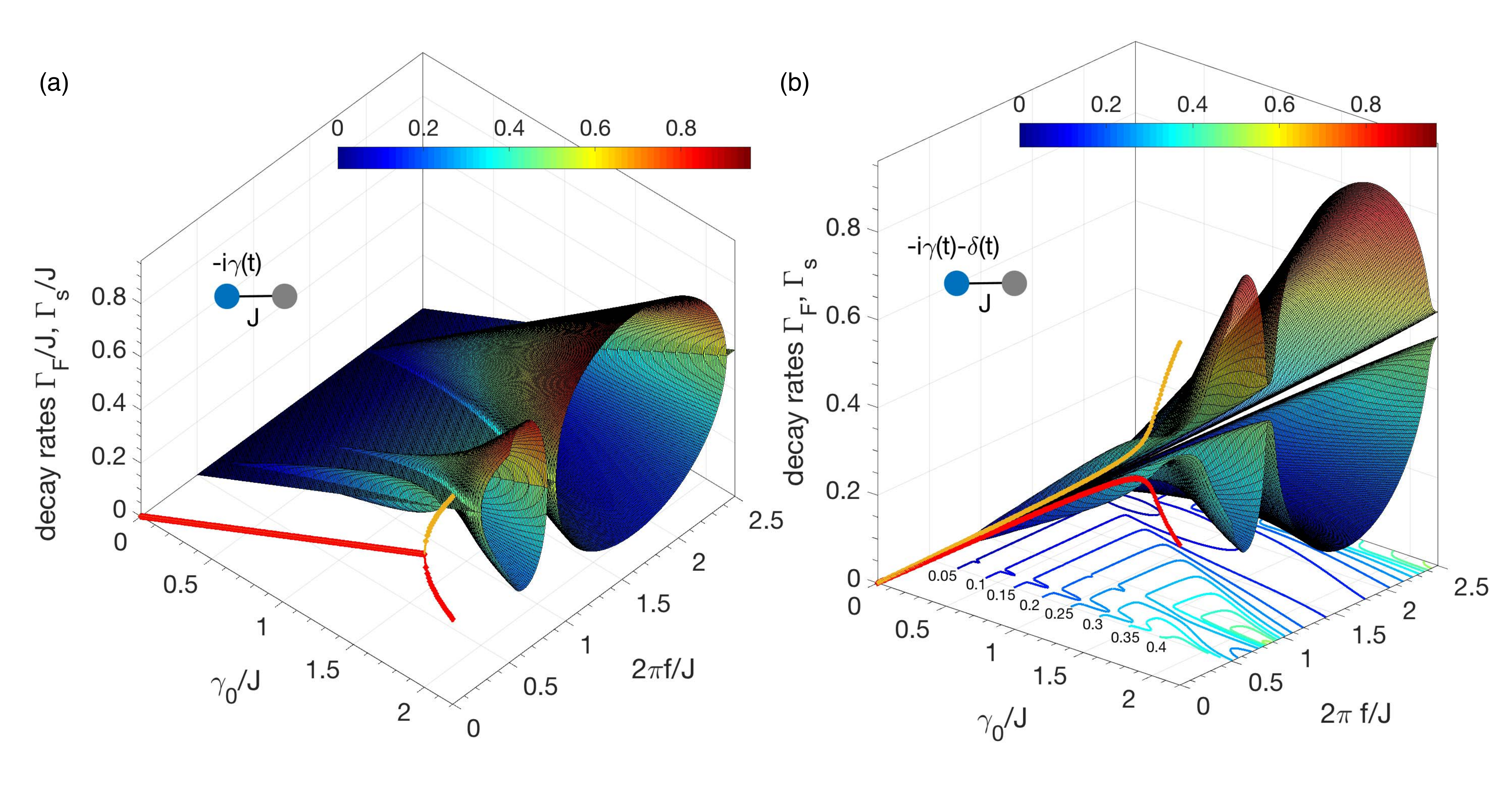}
\caption{Fast- and slow-mode decay rates $\Gamma_{F,s}$ for (a) symmetric, $\delta=0$, and (b) asymmetric, $\delta\neq 0$, dimer with effective tunneling amplitude $J$ and a Floquet loss. (a) for a symmetric dimer, the static result (filled, red/yellow circles) shows a classic bifurcation at the EP $\gamma_0/J=2$. The Floquet result shows the emergence of a slowly decaying mode at $\gamma_0/J\ll 1$ in the vicinity of resonances $2\pi f/J=2,2/3$. In this case, the $\mathcal{PT}$ transitions occur at an EP. (b) for an asymmetric dimer with $\delta/J=0.05$, the static result (filled, red/yellow circles) shows unequal decay rates with $d\Gamma/d\gamma>0$ for small $\gamma_0$, leading to slow-mode (filled red circles) for $\gamma_0/J\gtrsim 2$, i.e. at the location of the ALC. Floquet results show that a slow mode emerges at $\gamma_0/J\ll 1$ in the vicinity of $2\pi f/J=2,2/3$. In static and Floquet cases, the nonzero separation between the two decay rates shows that all passive $\mathcal{PT}$-symmetry breaking transitions in the asymmetric dimer occur at the location of the ALC.}
\label{fig:floquetEP}
\end{figure*}

Is the phenomenon of a passive $\mathcal{PT}$-symmetry breaking transition, which occurs without an EP and is driven by an ALC, a property singular to our system? Or is it broadly present in dissipative systems that are not ``identity shifted'' from a balanced gain and loss system? The answer to the latter question is a yes. In the weak coupling limit ($M\rightarrow 0,\omega_0\rightarrow\infty$) the electrical, two-oscillator system maps onto a dimer with tunneling amplitude $J=\omega_0M^2/2=\mathrm{const.}$ (Methods, section D). When the dissipation in the first oscillator is taken into account, the effective dimer Hamiltonian becomes
\begin{equation}
\label{eq:asymdimer}
H_{\mathrm{d}}(t)=-J\sigma_x-\frac{i\gamma_{\mathrm{d}}(t)+\delta}{2}({\bf 1}_2+\sigma_z).
\end{equation}
where $\sigma_k$ are the Pauli matrices, $\gamma_{\mathrm{d}}(t)=\gamma(t)/2$ is the Floquet loss in one level of the dimer, $\gamma(t)$ is the square-wave dissipation with a mean-value of $\gamma_0/2$, Eq.(\ref{eq:fl}), and the on-site-potential asymmetry $\delta$ is present only in the lossy circuit.

When $\delta=0$ Eq.(\ref{eq:asymdimer}) reduces to the classic case~\cite{Guo2009,ornigotti2014}. Figure~\ref{fig:floquetEP}a shows its decay rates $\Gamma_k$ as a function of loss strength $\gamma_0$ and modulation frequency $f$. In the static case, i.e. $f=0$, (filled red/yellow circles), the decay rates are equal to each other and increase with the loss strength when $\gamma_0/J<2$. The slowly decaying mode (filled red circles) emerges past the passive transition at the EP $\gamma_0=2J$. In the Floquet case, the surface plots for $0.5\leq 2\pi f/J\leq 2.5$ show that the passive $\mathcal{PT}$ transition occurs at vanishingly small $\gamma_0$ when the modulation frequency is near a resonance, i.e. $2\pi f/J=2,2/3,\cdots$~\cite{joglekar2014,lee2015,Li2016}. The lines of EPs that separate the fast-mode decay-rate surface $\Gamma_F(\gamma_0,f)$ and the slow-mode decay-rate surface $\Gamma_s(\gamma_0,f)$ are also visible.

Figure~\ref{fig:floquetEP}b shows the results for an asymmetric dimer with $\delta=0.05J$. In the static case (filled red/yellow circles), the two, slightly unequal decay rates increase with the loss strength, $d\Gamma_k/d\gamma>0$, when the loss strength is small. That changes for $\gamma/J\gtrsim 2$, where one eigenmode becomes slowly decaying, $d\Gamma_s/d\gamma<0$ (filled red circles), without an attendant EP. In the Floquet case, the surface plots for decay rates indicate the emergence of a slow mode at $\gamma_0/J\ll 1$ in the vicinity of resonances $2\pi f/J=2,2/3,\cdots$. However, the nonzero separation between the two surfaces clearly signals that the passive $\mathcal{PT}$ transitions occur at the location of the ALC. The contour lines of the slow-mode decay rate in the $(\gamma_0,f)$ plane also show that in the vicinity of resonances, $\Gamma_s$ becomes smaller with increasing loss strength $\gamma_0$. In our experiments with Floquet dissipation, the coupling between oscillators is $M=1/2\sqrt{2}=0.35$, and the oscillator frequency is $\omega_0=2\pi\times503$ s$^{-1}$; this gives the dimer tunneling amplitude $J=2\pi\times 30$ s$^{-1}$. Thus, the observed sequence of transitions in the vicinity of $f=60$ s$^{-1}$ in Fig.~\ref{fig:floquetloss}d corresponds to the primary resonance at $2\pi f/J=2$. Remarkably, $\mathcal{Q}(t)$ decay dynamics at moderate coupling shows emergence of slowly decaying eigenmodes at multiple frequencies that are not captured by the asymmetric dimer model.

Non-Hermitian degeneracies, exceptional points, and avoided level crossings play an important role in the dynamics of classical, gain-loss $\mathcal{PT}$ symmetric systems. Truly quantum versions of such  systems, however, are likely to be of a dissipative nature~\cite{xiao2017}, and may or may not be ``identity shifted'' from a balanced gain-loss system. Our results show that in such dissipative systems, the location of the ALC, where the eigenvalue flows are shortest distance apart, is instrumental to the passive $\mathcal{PT}$-symmetry breaking transition. With its versatility, our system provides a starting point for investigating the effects of interaction (nonlinearity), time-delay, and memory - all of which can be implemented via synthetic electronic circuits -- on the dynamics of dissipative $\mathcal{PT}$ symmetric systems.


\acknowledgments

This work was supported by CONACYT under the project CB-2016-01/284372, and by DGAPA-UNAM under the project UNAM-PAPIIT IA100718. MAQJ acknowledges CONACyT for a PhD scholarship and also thanks IPN for financial support through the grant SIP-IPN-20170023. RQT thanks financial support by the program UNAM-DGAPA-PAPIIT, Grant number IN112017. JLDJ thanks Catedras CONACYT-UNAM. JLA wishes to thank financial support from DGAPA-UNAM, under grant UNAM-PAPIIT IN110817. AKH and YNJ acknowledge financial support by NSF grant DMR 1054020.


\section*{Methods}


\subsection{Hamiltonian description from Kirchoff laws}
\label{sec:appA}

The equations of motion for the voltages $V_{1,2}(t)$ across the two capacitors $C$, the currents $I_{1,2}(t)$ across the two inductors $L$, and the current $I_x$ across the coupling inductor $L_x$ in Fig.~\ref{fig:static}a  are determined by Kirchoff laws, and are given by
\begin{eqnarray}
\label{eq:electronic}
\frac{dV_{1}}{dt} &=& -\frac{1}{RC}V_{1} - \frac{1}{C}I_{1} - \frac{1}{C}I_{x}, \nonumber \\
\frac{dV_{2}}{dt} &=& -\frac{1}{C}I_{2} + \frac{1}{C}I_{x}, \nonumber \\
\frac{dI_{1}}{dt} &=& \frac{1}{L}V_{1}, \\
\frac{dI_{2}}{dt} &=& \frac{1}{L}V_{2}, \nonumber \\
\frac{dI_{x}}{dt} &=& \frac{1}{L_{x}}(V_{1} -V_{2}). \nonumber
\end{eqnarray}
This set of five linear equations can be written in a matrix form, $i\partial_t|\phi(t)\rangle=\tilde{H}|\phi(t)\rangle$ where $|\phi\rangle=(V_1,V_2,I_1,I_2,I_x)^T$ is a real column vector, and the purely imaginary, non-symmetric, non-Hermitian matrix $\tilde{H}$ is
\begin{equation}
\label{Eq:Ham1}
\tilde{H} = i\left[
\begin{array}{ccccc}
-\frac{1}{RC} & 0 & -\frac{1}{C} & 0 & -\frac{1}{C}\\
0 & 0 & 0 & -\frac{1}{C} & \frac{1}{C}\\
\frac{1}{L} & 0 & 0 & 0 & 0 \\
0 & \frac{1}{L} & 0 & 0 & 0 \\
\frac{1}{L_{x}} & -\frac{1}{L_{x}} & 0 & 0 & 0\\
\end{array}\right].
\end{equation}
The energy in this circuit is given by
\begin{equation}
\label{eq:energy}
\mathcal{Q}(t)= \frac{1}{2}CV_{1}^{2}+\frac{1}{2}CV_{2}^{2}+\frac{1}{2}LI_{1}^{2} + \frac{1}{2}LI_{2}^{2} + \frac{1}{2}L_{x}I_{x}^{2}
\end{equation}
and can be represented by a positive quadratic form, i.e. $\mathcal{Q}=\langle\phi|A|\phi\rangle$ where $A=\mathrm{diag}(C,C,L,L,L_x)/2$ is a diagonal matrix. Defining a new variable $|\psi\rangle=A^{1/2}|\phi\rangle$ with the dimensions of square-root of energy ($\sqrt{\mathrm{Joule}}$), in the static case, the Kirchoff-law equations (\ref{eq:electronic}) lead to
\begin{eqnarray}
\label{eq:schr1}
i\partial_t|\psi(t)\rangle & = & H_D|\psi(t)\rangle,\\
\label{eq:schr2}
H_D & \equiv & A^{1/2}\tilde{H}A^{-1/2}.
\end{eqnarray}
Although $\tilde{H}$ is not Hermitian in the zero-loss case ($1/R=0$), the transformed Hamiltonian matrix $H_D$, Eq.(\ref{eq:Hd}), is Hermitian in that limit ($\gamma=0$). The dissipative Hamiltonian $H_D(\gamma)$ is shifted from its $\mathcal{PT}$ symmetric counterpart by $H_{PT}(\gamma/2)=H_{D}(\gamma/2)+i(\gamma/2){\bf I}_2$. The Hamiltonian $H_{PT}$ commutes with the $\mathcal{PT}$ operator where the block-diagonal, $5\times 5$ parity and time-reversal operators are given by
\begin{equation}
\label{eq:pt}
\mathcal{P}=\left(\begin{array}{ccc}
\sigma_x & 0 & 0\\
0 & \sigma_x & 0\\
0 & 0 & -1 \end{array}\right) ,
\hspace{3mm}
\mathcal{T}=\left(\begin{array}{cc}
{\bf 1}_2 & 0 \\ 0 & -{\bf 1}_3
\end{array}\right)\K,
\end{equation}
where ${\bf 1}_k$ is a $k\times k$ identity matrix, and $\K$ denotes complex conjugation.

When the circuit parameters are time dependent, the change-of-basis matrix $A^{1/2}(t)$ may become  time dependent as well. In this case, to change from the $|\phi\rangle$ basis to the $|\psi\rangle=A^{1/2}(t)|\phi\rangle$ basis, we have to include the gauge-field term that is generated by the time-dependent change of basis. Taking it into account gives
\begin{equation}
\label{eq:hdfloquet}
H_D(t)=\sqrt{A(t)}\tilde{H}(t)\frac{1}{\sqrt{A(t)}}-i\sqrt{A(t)}\partial_t \frac{1}{\sqrt{A(t)}}.
\end{equation}
In the Floquet dissipation case, the change-of-basis matrix is time independent and so the dissipative Hamiltonian $H_D(t)=A^{1/2}\tilde{H}(t)A^{-1/2}$ only has a time-dependent loss rate $\gamma(t)=1/R(t)C$. In the Floquet $L_x(t)$ case, the gauge-field term is nonzero and leads to
$H_D(t)=$
\begin{equation}
\label{eq:hdlx}
\left[
\begin{array}{ccccc}
-i\gamma & 0 & -i\omega_0 & 0 & -i\omega_0 M(t)\\
0 & 0 & 0 & -i\omega_0 & i\omega_0 M(t)\\
i\omega_0 & 0 & 0 & 0 & 0 \\
0 & i\omega_0 & 0 & 0 & 0 \\
i\omega_0 M(t)& -i\omega_0 M(t) & 0 & 0 & -i\partial_t\ln M(t)\\
\end{array} \right],
\end{equation}
where $M(t)=\sqrt{L/L_x(t)}$ characterizes the dimensionless coupling between the lossy oscillator and the neutral oscillator. Equations (\ref{eq:schr2}), (\ref{eq:hdfloquet}), and (\ref{eq:hdlx}) thus provide the requisite mapping from the Kirchoff-laws description to the Hamiltonian description.


\subsection{Circuit design and parameters}
\label{sec:appB}

Our experimental setup starts with two identical RLC electrical oscillators coupled by an inductor. The objectives of this work demands versatility in defining the elements used in the oscillators as well as the coupling, in terms of static and dynamic changes in magnitude. For the static condition, a controllable and fixed value is needed, and for the dynamic case, a precise control in the magnitude, frequency, and phase is required. Additionally, two independent and synchronized clocks are needed as well, one to define the initial conditions and the second to define the demanded dynamic changes. The solution is possible with the help of electronically synthesized circuits, using an analog computer built from different configurations of operational amplifiers~\cite{leon2015,roberto2014,quiroz2016}.

The problems associated with the faulty contacts and poor stability are resolved by mounting and soldering the electronic components of each oscillator and the coupling on a printed circuit board (PCB). The PCBs are designed in Altium Software and fabricated in a standard chemical etching process. Along these lines, the reproducibility and stability requires components to control offset, drift, and hidden frequency dependence. The implementation includes metal resistors (1\% tolerance), polyester capacitors, operational amplifiers (MC1458 and LF353) and analog multipliers (AD633). A stable DC power source is used to energize the electronic circuit, particularly, the 12 V bias voltage for the operational amplifiers.

The voltage signals in the electronic circuit correspond to the physical variables used in the mathematical model. The dynamics of the system are followed by measuring independently the variables of each oscillator (the voltage in the capacitor and the current in the inductor) and the coupling current, i.e. we measure the real, time-dependent vector $|\phi(t)\rangle$. The acquisition of the voltage signals is performed with a Rohde \& Schwarz oscilloscope, which has a 12-bits resolution in its analog/digital converter (impedance 1 M$\Omega$), and can transmit directly to a computer through a PC-OSCILLOSCOPE interface, which transfers the information by a USB connection. {\it Each measurement is averaged up to 64 times to reduce the influence of the electronic noise associated to the components.}

The initial input energy $\mathcal{Q}(t=0)$ is injected into the system using an Arbitrary Waveform Generator (AWG) from Agilent 33220A. The signal generated consists of a single pulse with a pulse duration of 0.2 ms and frequency 5 s$^{-1}$. The high-level voltage amplitude is 5 V, while the low-level voltage amplitude is 0 V. Besides setting the initial conditions, this clock synchronizes the system with the components' dynamical changes.

To implement the Floquet Hamiltonians, dynamic variations are introduced in the system by means of changes in the desired element of the system. The frequency, phase, and magnitude of the changes are defined by the period, phase, and amplitude of a voltage signal, which is provided by the second AWG from Agilent 33220A. The high- and the low-voltage amplitude levels correspond to the high and low energy-dissipation in the resistor, whereas in the case of dynamic coupling, the high level corresponds to a high inductance value, and the low level is related to a low inductance. All experiments are performed using the same PCB. The different configurations are reached by means of three mechanical selectors that remain in place during the course of the experiments.


\begin{figure}
\centering
\includegraphics[width=\columnwidth]{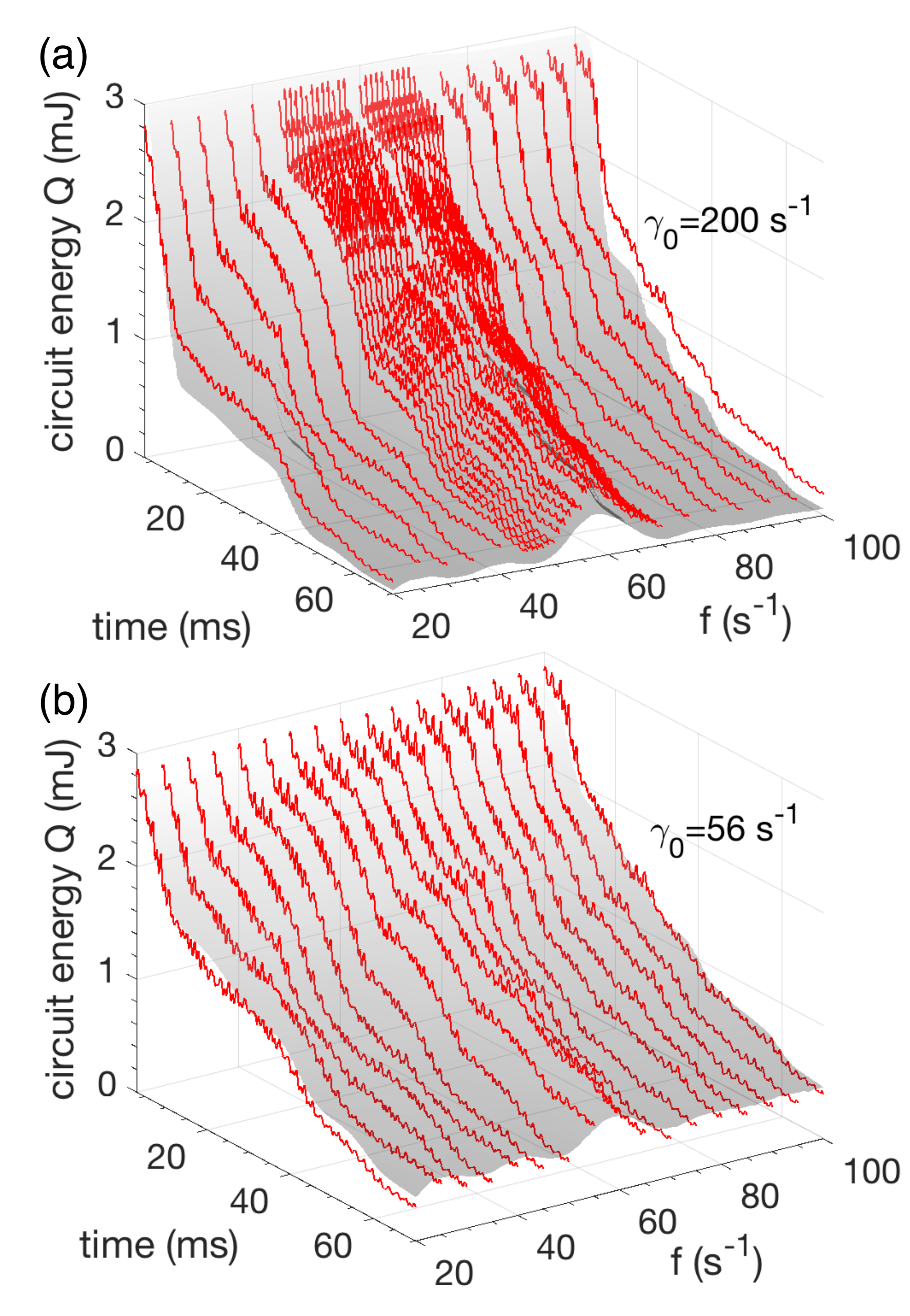}
\caption{Time-scans for the circuit energy $\mathcal{Q}(t,f)$ for different loss strengths. When $\gamma_0$=200 s$^{-1}$, (a), the slowly decaying mode in the vicinity of $f=60$ s$^{-1}$ is prominent, while for $\gamma_0$=56 s$^{-1}$, (b), it is visible. In all cases, the data (red lines) match the theory (surface) well.}
\label{fig:sampleQscans}
\end{figure}
\subsection{Quantitative analysis of agreement between theory and experiment}
\label{sec:appC}

In this section we provide a quantitative analysis of the similarity between our experimental results and the theoretical predictions. For this, first, we focus on the raw data of the experiment, that is, the decaying-energy $\mathcal{Q}(t)$ directly measured in the circuit, Eq.(\ref{eq:energy}). Figure~\ref{fig:sampleQscans} shows typical energy scans in the time-modulation frequency plane for two different $\gamma_0$ values (red lines: data; surface: theory). In each case, we see that the energy decay rate is dramatically lowered at $f=60$ s$^{-1}$ and the relative magnitude of the change is larger for higher loss strength $\gamma_0$.

\begin{figure}
\centering
\includegraphics[width=\columnwidth]{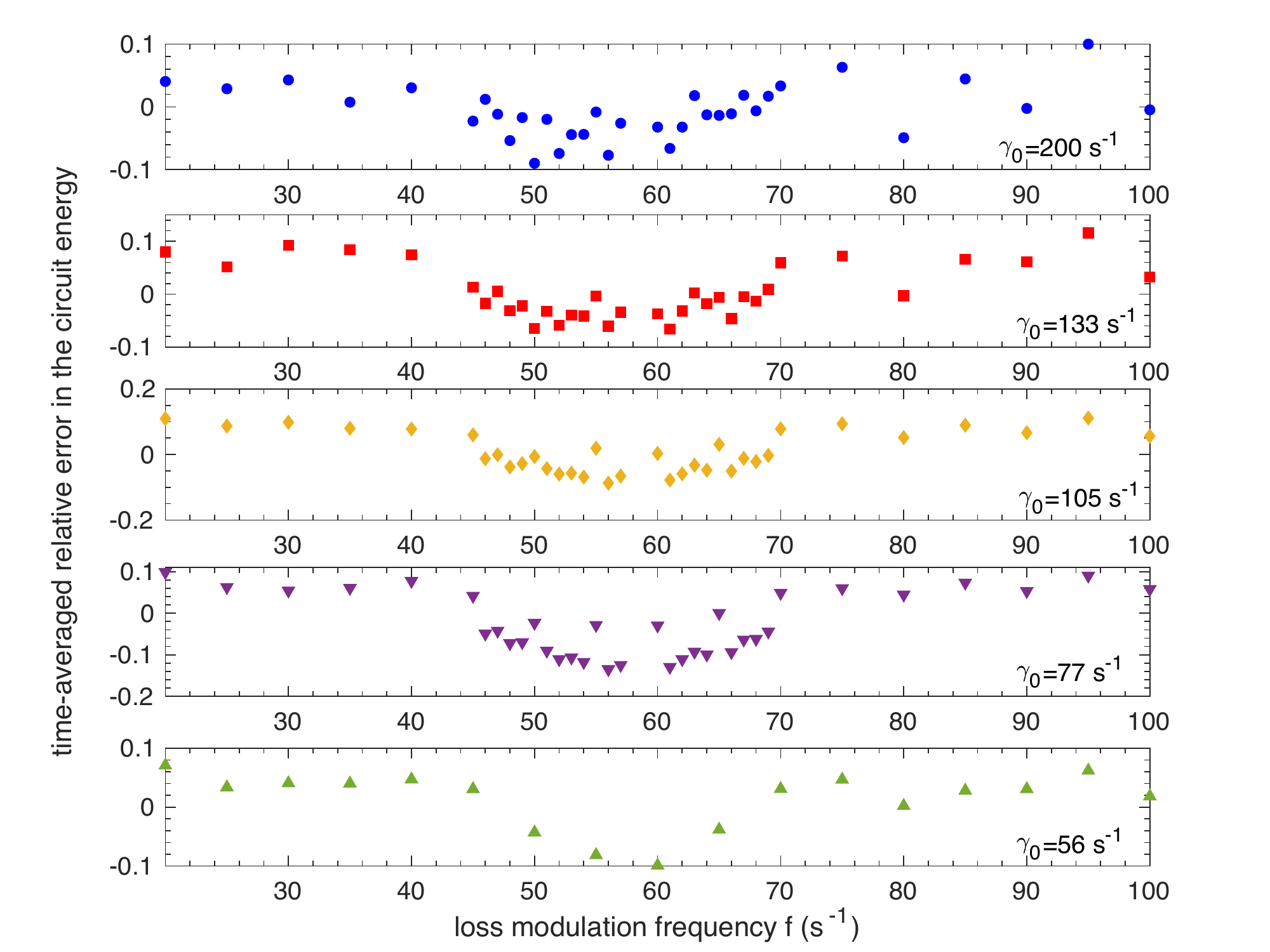}
\begin{tabular}{|c|c|c|}
\hline
loss rate $\gamma_0$ (s$^{-1}$) & relative $\delta\mathcal{Q}(\gamma_0)$ & relative $\delta\mu(\gamma_0)$ \\
\hline
200 & $-0.0075\pm0.042$ & $0.0019\pm0.0535$ \\\hline
133 & $0.053\pm0.051$ & $0.0046\pm0.0252$ \\\hline
105 & $0.0098\pm0.06$ & $0.0016\pm0.0176$ \\\hline
77 & $-0.025\pm0.076$ & $0.0025\pm0.0091$\\\hline
56 & $0.013\pm0.049$ & 0$\pm$0\\\hline
\end{tabular}
\caption{Quantifying the relative error in the circuit energy, $\delta\mathcal{Q}$, and the ratio, $\delta\mu$, for Floquet dissipation data. Plots of $\delta\mathcal{Q}(\gamma_0,f)$ as a function $f$ for different $\gamma_0$ show that the error is typically positive in the $\mathcal{PT}$ symmetric region and negative in the $\mathcal{PT}$ broken region. Table showing the frequency-averaged relative error in the circuit energy and the ratio quantifies the good agreement between theory and experimental results.}
\label{fig:relativeQerror}
\end{figure}

We define a time-averaged relative error for a given loss strength $\gamma_0$ and frequency $f$ as
\begin{equation}
\label{eq:deltaQ}
\delta\mathcal{Q}(\gamma_0,f)\equiv 1-\frac{1}{2\tau}\int_0^{2\tau}\frac{\mathcal{Q}^{\mathrm{exp}}(t')}{\mathcal{Q}^{\mathrm{th}}(t')}dt',
\end{equation}
where $\mathcal{Q}^{\mathrm{exp}}(t')$ is the experimentally measured circuit energy and $\mathcal{Q}^{\mathrm{th}}(t')$ is the theoretical prediction for it. The resulting relative error values for the Floquet-loss experimental data are shown in Fig.~\ref{fig:relativeQerror}. We also define the relative error in the ratio, Eq.(\ref{eq:ratio}), as
\begin{equation}
\label{eq:deltamu}
\delta\mu(\gamma_0)=1-\left\langle\frac{\mu^{\mathrm{exp}}(\gamma_0,f)}{\mu^{\mathrm{th}}(\gamma_0,f)}\right\rangle_f
\end{equation}
where $\langle\cdots\rangle_f$ denotes the average over loss-modulation frequencies. The second and third column in the table show the frequency-averaged $\delta\mathcal{Q}(\gamma_0)$ and $\delta\mu(\gamma_0)$ for the experimental data.

\begin{figure}
\centering
 \includegraphics[width=\columnwidth]{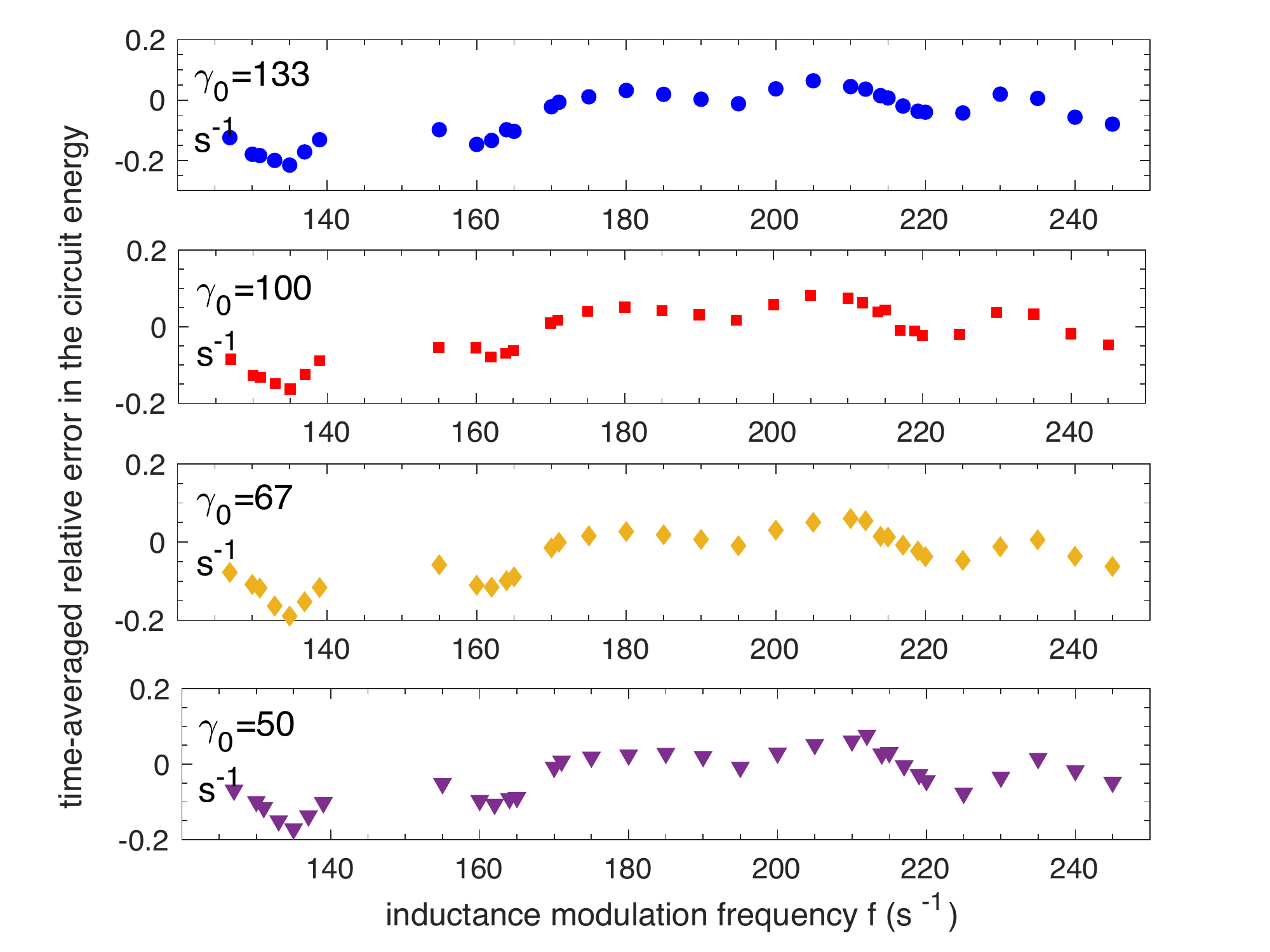}
\begin{tabular}{|c|c|c|}
\hline
loss rate $\gamma_0$ (s$^{-1}$) & relative $\delta\mathcal{Q}(\gamma_0)$ & relative $\delta\mu(\gamma_0)$ \\
\hline
133 & $0.0085\pm0.023 $ & $-0.055\pm0.081$ \\\hline
100 & $0.01\pm0.029$ & $-0.021\pm0.07$ \\\hline
67 & $0.0068\pm0.023$ & $-0.041\pm0.067$\\\hline
50 & $0.0049\pm0.033$ & -0.035$\pm0.066$\\\hline
\end{tabular}
\caption{Relative error in the circuit energy $\delta\mathcal{Q}(\gamma,f)$ shows that $\mu>0$ regions correlate with negative $\delta\mathcal{Q}$, as they do in the Floquet dissipation case, Fig.~\ref{fig:relativeQerror}. The time-window used for calculating the ratio $\mu$ is $\tau$=7 ms. Table showing the frequency-averaged relative error in the circuit energy and the ratio quantifies the  agreement between theory and experimental results.}
\label{fig:relativeQerrorLx}
\end{figure}

Figure~\ref{fig:relativeQerrorLx} shows that the relative error in the circuit energy $\delta\mathcal{Q}(\gamma,f)$ for the dynamic coupling case is, typically, larger than that in the dynamic dissipation case. This is due to the fact that any change in the coupling between oscillators, represented by the inductor $L_{x}$, removes (injects) energy from (into) the system. This creates instabilities in the experimental system, which leads to a larger uncertainty in the measurement of the circuit variables.

\subsection{Equivalence between quantum and electrical-oscillator systems}
\label{sec:appD}

The dynamics of a single excitation in a system comprising two coupled quantum oscillators is described by the Schr\"{o}dinger equation
\begin{equation}
\label{Eq:quantum1}
i\partial_t\ket{\psi\pare{t}} = \hat{H}_{\mathrm{osc}}\ket{\psi\pare{t}},
\end{equation}
where the Hamiltonian $\hat{H}_{\mathrm{osc}}$ is given by
\begin{equation}
\hat{H}_{\mathrm{osc}} = \sum_{n=1}^{2} \varepsilon_{n}\ket{n}\bra{n} + \sum_{n \neq m}^{2}J_{nm}\ket{n}\bra{m},
\end{equation}
with $\ket{n}$ denoting the energy density at the $n$th oscillator. The $n$th-site energies and the coupling between sites $n$ and $m$ are described by $\varepsilon_{n}$ and $J_{nm}$, respectively. By expanding the time-dependent wavefunction in the site basis, i.e. $\ket{\psi\pare{t}} = \sum_{n}c_{n}\pare{t}\ket{n}$, it is easy to find that Eq. (\ref{Eq:quantum1}) leads to a set of coupled equations of first order in the time derivative,
\begin{equation}\label{Eq:quantum2}
i\partial_t c_{n} = \varepsilon_{n}c_{n}\pare{t} + \sum_{n \neq m}^{2}J_{nm}c_{m}\pare{t}.
\end{equation}
In the weak-coupling limit ($J_{nm} \ll \varepsilon_{n}$), the time-derivative of Eq. (\ref{Eq:quantum2}) becomes~\cite{briggs2011,roberto2013}
\begin{equation}\label{Eq:quantum3}
\partial^2_t c_{n} + \varepsilon_{n}^{2}c_{n} + \varepsilon_{n}\sum_{n \neq m}^2 2J_{nm}c_{m}
\end{equation}
Thus, by considering similar oscillators ($\varepsilon=\varepsilon_{1}\simeq\varepsilon_{2}$), we can define $K=2\varepsilon J_{12}=2\varepsilon J_{21}=2\varepsilon J$ and obtain~\cite{briggs2011}
\begin{eqnarray}
\partial_t^2 c_{+} + \pare{\varepsilon^{2} + K}c_{+} &=& 0, \label{Eq:c+}\\
\partial_t^2 c_{-} + \pare{\varepsilon^{2} - K}c_{-} &=& 0, \label{Eq:c-}
\end{eqnarray}
where $c_{\pm}=c_{1} \pm c_{2}$ denote the normal modes of the two oscillator system.

To establish a connection between the quantum model and our experimental setup, let us consider Eq. (\ref{eq:electronic}) in the non-dissipative limit, that is, when $1/R=0$,
\begin{eqnarray}
\frac{dV_{1}}{dt} &=& - \frac{1}{C}I_{1} - \frac{1}{C}I_{x}, \nonumber \\
\frac{dV_{2}}{dt} &=& -\frac{1}{C}I_{2} + \frac{1}{C}I_{x}, \nonumber \\
\frac{dI_{1}}{dt} &=& \frac{1}{L}V_{1}, \\
\frac{dI_{2}}{dt} &=& \frac{1}{L}V_{2}, \nonumber \\
\frac{dI_{x}}{dt} &=& \frac{1}{L_{x}}V_{1} - \frac{1}{L_{x}}V_{2}. \nonumber
\end{eqnarray}
It is straightforward to rewrite these equations as~\cite{quiroz2016}
\begin{eqnarray}
\partial_t^2V_{+} + \pare{\omega_0^2(1+M^2)+\omega_0^2M^2 }V_{+} &=& 0, \label{Eq:V+}\\
\partial_t^2V_{-}+ \pare{\omega_0^2(1+M^2)-\omega_0^2M^2 }V_{-} &=& 0, \label{Eq:V-}
\end{eqnarray}
where $V_{\pm}=V_{1} \pm V_{2}$ are the symmetric and antisymmetric normal modes of two LC circuits, $\omega_0=1/\sqrt{LC}$ is the frequency of an isolated LC circuit, and $M^2=L/L_x$.

By comparing Eqs. (\ref{Eq:c+})-(\ref{Eq:c-}) with Eqs. (\ref{Eq:V+})-(\ref{Eq:V-}), we find that our experimental setup, in the weak-coupling regime, is equivalent to a quantum-mechanical system by
setting $c_{1,2}\rightarrow V_{1,2}$, $\varepsilon\rightarrow\omega_0\sqrt{1+M^2}$, and an effective tunneling amplitude
\begin{equation}
J\rightarrow \frac{\omega_0M^2}{2\sqrt{1+M^2}}.
\end{equation}
Thus, the weak-coupling limit can be formally defined by $M\rightarrow 0$, $\omega_0\rightarrow\infty$ such that the product $J=\omega_0 M^2/2$ remains constant.

%

%


\begin{thebibliography}{XX}
\bibitem{caveat} We use this criterion for a maximum, instead of $d^2\Gamma_s/d\gamma^2=0$, because the derivative $d\Gamma_s/d\gamma$ is not necessarily differentiable; in the case of an exceptional point, it has a kink.
\end{thebibliography}
\end{document}